\documentclass[english,DIV=calc, paper=a4, fontsize=11pt]{scrartcl}
\usepackage[T1]{fontenc}
\usepackage[latin9]{inputenc}
\usepackage{geometry}
\usepackage{color}
\usepackage{array}
\usepackage{float}
\usepackage{amsmath}

\makeatletter

\providecommand{\tabularnewline}{\\}

\usepackage[english]{babel} % English language/hyphenation
\usepackage{amsmath,amsfonts,amsthm,amssymb,amsbsy,mathtools} % Math packages
\usepackage{palatino} % in LyX, standard font isn't vector graphics
\usepackage{multicol}
\usepackage[svgnames]{xcolor} % Enabling colors by their 'svgnames'
\usepackage[hang, small,labelfont=bf,up,textfont=it,up]{caption} % Custom captions under/above floats in tables or figures
\usepackage{booktabs} % Horizontal rules in tables
\usepackage{fix-cm}	 % Custom font sizes - used for the initial letter in the document
\usepackage{graphicx}	
\usepackage{sectsty} % Enables custom section titles
\usepackage{ulem}
\usepackage{soul} % for strikethrough \st
\allsectionsfont{\usefont{OT1}{phv}{b}{n}} % Change the font of all section commands
\usepackage{cite}
\usepackage{fancyhdr} % Needed to define custom headers/footers
\usepackage{lastpage} % Used to determine the number of pages in the document (for "Page X of Total")
\usepackage[breaklinks=true]{hyperref}
\usepackage{doi}
\hypersetup{
  colorlinks   = true, %Colours links instead of ugly boxes
  urlcolor     = blue, %Colour for external hyperlinks
  linkcolor    = blue, %Colour of internal links
  citecolor   = red %Colour of citations
}
\usepackage{url}

\usepackage{pifont}
\newcommand{\cmark}{\ding{51}}%
\newcommand{\xmark}{\ding{55}}%

\newcommand\SmallMatrix[1]{{%
\arraycolsep=0.85\arraycolsep\ensuremath{\begin{pmatrix}#1\end{pmatrix}}}}

\usepackage{lettrine} % Package to accentuate the first letter of the text
\newcommand{\initial}[1]{ % Defines the command and style for the first letter
\lettrine[lines=3,lhang=0.3,nindent=0em]{
\color{DarkGoldenrod}
{\textsf{#1}}}{}}

\renewcommand\emph[1]{{\it{#1}}}

\usepackage{titling} % Allows custom title configuration

\newcommand{\HorRule}{\color{DarkGoldenrod} \rule{\linewidth}{1pt}} % Defines the gold horizontal rule around the title

\pretitle{\vspace{-80pt} \begin{flushleft} \HorRule \fontsize{24}{24} \usefont{OT1}{phv}{b}{n} \color{DarkRed} \selectfont} % Horizontal rule before the title

\title{Weak-value amplification: state of play} % Your article title

\posttitle{\par\end{flushleft}\vskip 0.5em\vspace{-20pt}} % Whitespace under the title

\posttitle{\par\end{flushleft}\vskip 0.5em\vspace{-20pt}} % Whitespace under the title

\preauthor{\begin{flushleft}\large \lineskip 0.5em \usefont{OT1}{phv}{b}{sl} \color{DarkRed}} % Author font configuration

\postauthor{\footnotesize \usefont{OT1}{phv}{m}{sl} \color{Black} % Configuration for the institution name
\newline $^1$Quantum Chemistry Laboratory, Kyoto University,  606-8502, Kyoto, Japan, $^2$Physical \& Theoretical Chemistry Laboratory, Oxford University, OX1 3QZ, Oxford, UK, $^3$University of Calgary, T2N 4N1, Calgary, Canada.   $^*$\href{mailto:dattani.nike@gmail.com}{dattani.nike@gmail.com}, $^\dagger$\href{mailto:nbryans1@gmail.com}{nbryans1@gmail.com}

\vspace{-20pt}
\par\end{flushleft}\HorRule} % Horizontal rule after the title

\makeatother

\usepackage{babel}
\begin{document}

\title{{\fontsize{0.72cm}{1em}\selectfont Quantum factorization of 56153 with only 4 qubits}}

\author{Nikesh S. Dattani,$^{1,2,*}$ Nathaniel Bryans$^{\,3,\dagger}$}
%	\date{} % no date

\maketitle
\vspace{-7mm}

\begin{multicols}{2}

\vspace{-30mm}
\initial{T}\textbf{he largest number factored on a quantum device
reported until now was 143 \cite{Xu2012}. That quantum computation,
which used only 4 qubits at 300K, actually also factored much larger numbers
such as 3599, 11663, and 56153, without the awareness of the authors
of that work. Furthermore, unlike the implementations of Shor's algorithm
performed thus far \cite{Vandersypen2001,Lanyon2007,Lu2007,Politi2009,Martin-Lopez2012,Lucero2012,Smolin2013},
these 4-qubit factorizations do not need to use prior knowledge of
the answer. However, because they only use 4 qubits, these factorizations
can also be performed trivially on classical computers. We discover
a class of numbers for which the power of quantum information actually
comes into play. We then demonstrate a 3-qubit} \textbf{factorization
of 175, which would be the first quantum factorization of a triprime.}
\vspace{0.5mm}

It is well known that factoring large numbers on classical computers
is extremely resource demanding, and that Shor's algorithm could theoretically
allow a quantum computer to factor the same numbers with drastically
fewer operations. However, in its 20-year lifespan, Shor's algorithm
has not gone far in terms of factoring large numbers. Until 2012
the largest number factored using Shor's algorithm was 15 \cite{Vandersypen2001,Lanyon2007,Lu2007,Politi2009,Martin-Lopez2012,Smolin2013},
and today the largest is still only 21 \cite{Lucero2012,Smolin2013}.
Furthermore, these factorizations were not genuine implementations
of Shor's algorithm because they relied on prior knowledge of the
answer to the factorization problem being solved in the first place
\cite{Smolin2013}. 

An alternative to Shor's algorithm that also makes use of quantum
mechanics to factor numbers, relies on first transforming the factorization
problem into an optimization problem. This idea was first introduced
in 2001 by Burges \cite{Burges2002}, and it was improved in 2010
by Schaller and Schutzhold \cite{Schaller2010}, and then again in
2012 by Xu \emph{et al} \cite{Xu2012} who used it with NMR to factor
the number 143 with 4 qubits and no prior knowledge of the answer
to the problem. To provide context, factoring the number 15 with Shor's
algorithm and no prior knowledge of the answer to the problem, requires
a minimum of 8 qubits (and more if error correction is attempted);
and this has still never been done.\vspace{-3mm}

\section*{Quantum factorization of 143}

\vspace{-3mm}

The NMR factorization of 143 in 2012 \cite{Xu2012} began with the
multiplication table:

\vspace{-5mm}

\begin{table}[H]
\protect\caption{Multiplication table for $11\times13=143$ in binary.}

{\footnotesize{}}%
\begin{tabular}{c>{\centering}p{3.8mm}>{\centering}p{3.8mm}>{\centering}p{3.8mm}>{\centering}p{3.8mm}>{\centering}p{3.8mm}>{\centering}p{3.8mm}>{\centering}p{2.6mm}>{\centering}p{2.5mm}}
\hline 
 & \textbf{\footnotesize{}$2^{7}$} & \textbf{\footnotesize{}$2^{6}$} & \textbf{\footnotesize{}$2^{5}$} & \textbf{\footnotesize{}$2^{4}$} & \textbf{\footnotesize{}$2^{3}$} & \textbf{\footnotesize{}$2^{2}$} & \textbf{\footnotesize{}$2^{1}$} & \textbf{\footnotesize{}$2^{0}$}\tabularnewline
\hline 
\hline 
{\footnotesize{}$p$} &  &  &  &  & {\footnotesize{}1} & {\footnotesize{}$p_{2}$} & {\footnotesize{}$p_{1}$} & {\footnotesize{}1}\tabularnewline
{\footnotesize{}$q$} &  &  &  &  & {\footnotesize{}1} & {\footnotesize{}$q_{2}$} & {\footnotesize{}$q_{1}$} & {\footnotesize{}1}\tabularnewline
\hline 
{\footnotesize{}} &  &  &  &  & {\footnotesize{}1} & {\footnotesize{}$p_{2}$} & {\footnotesize{}$p_{1}$} & {\footnotesize{}1}\tabularnewline
 &  &  &  & {\footnotesize{}$q_{1}$} & {\footnotesize{}$p_{2}q_{1}$} & {\footnotesize{}$p_{1}q_{1}$} & {\footnotesize{}$q_{1}$} & \tabularnewline
 &  &  & {\footnotesize{}$q_{2}$} & {\footnotesize{}$p_{2}q_{2}$} & {\footnotesize{}$p_{1}q_{2}$} & {\footnotesize{}$q_{2}$} &  & \tabularnewline
 &  & {\footnotesize{}1} & {\footnotesize{}$p_{2}$} & {\footnotesize{}$p_{1}$} & {\footnotesize{}1} &  &  & \tabularnewline
{\footnotesize{}carries} & $z_{67}$ & $z_{56}$ & $z_{45}$ & $z_{34}$ & $z_{23}$ & $z_{12}$ &  & \tabularnewline
 & $z_{57}$ & $z_{46}$ & $z_{35}$ & $z_{24}$ &  &  &  & \tabularnewline
\hline 
{\footnotesize{}$p\times q=143$} & {\footnotesize{}1} & {\footnotesize{}0} & {\footnotesize{}0} & {\footnotesize{}0} & {\footnotesize{}1} & {\footnotesize{}1} & {\footnotesize{}1} & {\footnotesize{}1}\tabularnewline
\hline 
\end{tabular}{\footnotesize \par}

\end{table}

\vspace{-3mm}

Adding each column leads to the equations:

\vspace{-6mm}
{\small{}
\begin{eqnarray}
p_{1}+q_{1} & = & 1+2z_{12}\\
p_{2}+p_{1}q_{1}+q_{2}+z_{12} & = & 1+2z_{23}+4z_{24}\\
1+p_{2}q_{1}+p_{1}q_{2}+1+z_{23} & = & 1+2z_{34}+4z_{35}\\
 & \vdots\nonumber \\
q_{2}+p_{2}+z_{45}+z_{35} & = & 0+2z_{56}+4z_{57}\\
\mbox{\ensuremath{1+z_{56}+z_{46}}} & = & \mbox{\ensuremath{0+2z_{67}}}\\
\mbox{\ensuremath{z_{67}+z_{57}}} & = & \mbox{\ensuremath{1}.}
\end{eqnarray}
}{\small \par}

\vspace{-1mm}

By noticing simplifications such as $z_{12}=0$ when $p_{1}+q_{1}=1+2z_{12}$,
these equations then reduce to:\vspace{-4mm}

\begin{eqnarray}
p_{1}+q_{1}-1 & = & 0\label{eq:XuEquations1}\\
p_{2}+q_{2}-1 & = & 0\label{eq:XuEquations2}\\
p_{2}q_{1}+p_{1}q_{2}-1 & = & 0.\label{eq:XuEquations3}
\end{eqnarray}

%\vspace{-2mm}

Since any real number squared is $\ge0$, the minimum of the function
$(p_{1}+q_{1}-1)^{2}$ is 0, and the values of $(p_{1},q_{1})$ that
allow this minimum to be attained then must also be the solution to
equation \ref{eq:XuEquations1}. In fact, the minimum of the function:

\vspace{-6mm}

\begin{equation}
(p_{1}+q_{1}-1)^{2}+(p_{2}+q_{2}-1)^{2}+(p_{2}q_{1}+p_{1}q_{2}-1)^{2}\label{eq:XuObjectiveFunction}
\end{equation}

\vspace{-1mm}

\noindent is also 0, meaning that the values of $(p_{1},p_{2},q_{1},q_{2})$
that minimize Eq.\ref{eq:XuObjectiveFunction} are also the solution
to the equations \ref{eq:XuEquations1}-\ref{eq:XuEquations3}. This
means that the values of $(p_{1},p_{2},q_{1},q_{2})$ which represent
the solution to the factorization problem, are encoded in the ground
state of the 4-qubit Hamiltonian:

\end{multicols}

\vspace{2mm}

\noindent {\small{}}
\begin{table*}
{\small{}\protect\caption{Mutiplication table for $241\times233=56153$ in binary \label{tab:MutiplicationFor} }
}{\small \par}

\centering{}%
\begin{tabular}{>{\centering}p{6mm}>{\centering}p{6.5mm}>{\centering}p{6.4mm}>{\centering}p{6.4mm}>{\centering}p{6.3mm}>{\centering}p{6.3mm}>{\centering}p{6mm}>{\centering}p{6mm}>{\centering}p{6mm}>{\centering}p{6mm}>{\centering}p{6mm}>{\centering}p{6mm}>{\centering}p{6mm}>{\centering}p{6mm}>{\centering}p{5.7mm}>{\centering}p{3.2mm}>{\centering}p{2.3mm}}
\hline 
\noalign{\vskip0.9mm}
 & $2^{15}$ & $2^{14}$ & $2^{13}$ & $\textbf{\ensuremath{2^{12}}}$ & \textbf{$2^{11}$} & \textbf{$2^{10}$} & \textbf{$2^{9}$} & \textbf{$2^{8}$} & \textbf{$2^{7}$} & \textbf{$2^{6}$} & \textbf{$2^{5}$} & \textbf{$2^{4}$} & \textbf{$2^{3}$} & \textbf{$2^{2}$} & \textbf{$2^{1}$} & \textbf{$2^{0}$}\tabularnewline[0.5mm]
\hline 
\hline 
$p$ &  &  &  &  &  &  &  &  & 1 & $p_{6}$ & $p_{5}$ & $p_{4}$ & $p_{3}$ & $p_{2}$ & $p_{1}$ & 1\tabularnewline
$q$ &  &  &  &  &  &  &  &  & 1 & $q_{6}$ & $q_{5}$ & $q_{4}$ & $q_{3}$ & $q_{2}$ & $q_{1}$ & 1\tabularnewline
\hline 
\noalign{\vskip1mm}
 &  &  &  &  &  &  &  &  & 1 & $p_{6}$ & $p_{5}$ & $p_{4}$ & $p_{3}$ & $p_{2}$ & $p_{1}$ & 1\tabularnewline[0.8mm]
 &  &  &  &  &  &  &  & $q_{1}$ & $p_{6}q_{1}$ & $p_{5}q_{1}$ & $p_{4}q_{1}$ & $p_{3}q_{1}$ & $p_{2}q_{1}$ & $p_{1}q_{1}$ & $q_{1}$ & \tabularnewline
 &  &  &  &  &  &  & $q_{2}$ & $p_{6}q_{2}$ & $p_{5}q_{2}$ & $p_{4}q_{2}$ & $p_{3}q_{2}$ & $p_{2}q_{2}$ & $p_{1}q_{2}$ & $q_{2}$ &  & \tabularnewline
 &  &  &  &  &  & $q_{3}$ & $p_{6}q_{3}$ & $p_{5}q_{3}$ & $p_{4}q_{3}$ & $p_{3}q_{3}$ & $p_{2}q_{3}$ & $p_{1}q_{3}$ & $q_{3}$ &  &  & \tabularnewline
 &  &  &  &  & $q_{4}$ & $p_{6}q_{4}$ & $p_{5}q_{4}$ & $p_{4}q_{4}$ & $p_{3}q_{4}$ & $p_{2}q_{4}$ & $p_{1}q_{4}$ & $q_{4}$ &  &  &  & \tabularnewline
 &  &  &  & $q_{5}$ & $p_{6}q_{5}$ & $p_{5}q_{5}$ & $p_{4}q_{5}$ & $p_{3}q_{5}$ & $p_{2}q_{5}$ & $p_{1}q_{5}$ & $q_{5}$ &  &  &  &  & \tabularnewline
 &  &  & $q_{6}$ & $p_{6}q_{6}$ & $p_{5}q_{6}$ & $p_{4}q_{6}$ & $p_{3}q_{6}$ & $p_{2}q_{6}$ & $p_{1}q_{6}$ & $q_{6}$ &  &  &  &  &  & \tabularnewline
 &  & 1 & $p_{6}$ & $p_{5}$ & $p_{4}$ & $p_{3}$ & $p_{2}$ & $p_{1}$ & 1 &  &  &  &  &  &  & \tabularnewline
 & $z_{14,15}$ & $z_{13,14}$ & $z_{12,13}$ & $z_{11,12}$ & $z_{10,11}$ & $z_{9,10}$ & $z_{8,9}$ & $z_{7,8}$ & $z_{6,7}$ & $z_{5,6}$ & $z_{4,5}$ & $z_{3,4}$ & $z_{2,3}$ & $z_{1,2}$ &  & \tabularnewline
 & $z_{13,15}$ & $z_{12,14}$ & $z_{11,13}$ & $z_{10,12}$ & $z_{9,11}$ & $z_{8,10}$ & $z_{7,9}$ & $z_{6,8}$ & $z_{5,7}$ & $z_{4,6}$ & $z_{3,5}$ & $z_{2,4}$ &  &  &  & \tabularnewline
 &  & $z_{11,14}$ & $z_{10,13}$ & $z_{9,12}$ & $z_{8,11}$ & $z_{7,10}$ & $z_{6,9}$ & $z_{5,8}$ & $z_{4,7}$ &  &  &  &  &  &  & \tabularnewline[0.6mm]
\hline 
\noalign{\vskip1mm}
 & 1 & 1 & 0 & 1 & 1 & 0 & 1 & 1 & 0 & 1 & 0 & 1 & 1 & 0 & 0 & 1\tabularnewline[0.6mm]
\hline 
\end{tabular}
\end{table*}

\vspace{-5mm}
\vspace{-9mm}

\begin{eqnarray}
H & = & (p_{1}+q_{1}-1)^{2}+(p_{2}+q_{2}-1)^{2}+(p_{2}q_{1}+p_{1}q_{2}-1)^{2},\\
 & = & 5-3p_{1}-p_{2}-q_{1}+2p_{1}q_{1}-3p_{2}q_{1}+2p_{1}p_{2}q_{1}-3q_{2}+p_{1}q_{2}+2p_{2}q_{2}+2p_{2}q_{1}q_{2}\label{eq:4qubitHamiltonian}\\
a_{i} & = & \frac{1}{2}\left(1-\sigma_{z}^{(i)}\right)^{2}.
\end{eqnarray}

\begin{multicols}{2}

\vspace{-10mm}

It is easy to see that this is true, especially when we look at the
Hamiltonian in matrix form:

\vspace{-5mm}

\setcounter{MaxMatrixCols}{16}
\begin{flalign*}
\scriptstyle
\SmallMatrix{ 5 & 0 & 0 & 0 & 0 & 0 & 0 & 0 & 0 & 0 & 0 & 0 & 0 & 0 & 0 & 0\\0 & 2 & 0 & 0 & 0 & 0 & 0 & 0 & 0 & 0 & 0 & 0 & 0 & 0 & 0 & 0\\0 & 0 & 4 & 0 & 0 & 0 & 0 & 0 & 0 & 0 & 0 & 0 & 0 & 0 & 0 & 0\\0 & 0 & 0 & 1 & 0 & 0 & 0 & 0 & 0 & 0 & 0 & 0 & 0 & 0 & 0 & 0\\0 & 0 & 0 & 0 & 4 & 0 & 0 & 0 & 0 & 0 & 0 & 0 & 0 & 0 & 0 & 0\\0 & 0 & 0 & 0 & 0 & 3 & 0 & 0 & 0 & 0 & 0 & 0 & 0 & 0 & 0 & 0\\0 & 0 & 0 & 0 & 0 & 0 & \mathbf{0} & 0 & 0 & 0 & 0 & 0 & 0 & 0 & 0 & 0\\0 & 0 & 0 & 0 & 0 & 0 & 0 & 1 & 0 & 0 & 0 & 0 & 0 & 0 & 0 & 0\\0 & 0 & 0 & 0 & 0 & 0 & 0 & 0 & 2 & 0 & 0 & 0 & 0 & 0 & 0 & 0\\0 & 0 & 0 & 0 & 0 & 0 & 0 & 0 & 0 & \mathbf{0} & 0 & 0 & 0 & 0 & 0 & 0\\0 & 0 & 0 & 0 & 0 & 0 & 0 & 0 & 0 & 0 & 3 & 0 & 0 & 0 & 0 & 0\\0 & 0 & 0 & 0 & 0 & 0 & 0 & 0 & 0 & 0 & 0 & 1 & 0 & 0 & 0 & 0\\0 & 0 & 0 & 0 & 0 & 0 & 0 & 0 & 0 & 0 & 0 & 0 & 1 & 0 & 0 & 0\\0 & 0 & 0 & 0 & 0 & 0 & 0 & 0 & 0 & 0 & 0 & 0 & 0 & 1 & 0 & 0\\0 & 0 & 0 & 0 & 0 & 0 & 0 & 0 & 0 & 0 & 0 & 0 & 0 & 0 & 1 & 0\\0 & 0 & 0 & 0 & 0 & 0 & 0 & 0 & 0 & 0 & 0 & 0 & 0 & 0 & 0 & 3}.
\end{flalign*}

%\vspace{-2mm}

Since this matrix is diagonal, we can read off the eigenvalues, the
lowest of which are clearly at eigenstates $|6\rangle=|0110\rangle$
and $|9\rangle=|1001\rangle$, corresponding (respectively) to:

\vspace{-2mm}

\begin{eqnarray}
(p_{1},p_{2},q_{1},q_{2}) & \negmedspace\negmedspace\negmedspace=\negmedspace\negmedspace\negmedspace & (0,1,1,0)\rightarrow p=13,\, q=11,\qquad\\
(p_{1},p_{2},q_{1},q_{2}) & \negmedspace\negmedspace\negmedspace=\negmedspace\negmedspace\negmedspace & (1,0,0,1)\rightarrow p=11,\, q=13.\qquad
\end{eqnarray}

\vspace{-2mm}

\begin{table*}
\protect\caption{Multiplication table for $557\times523=291311$ in binary.\label{tab:10x10291311}}

\centering{}{\scriptsize{}}%
\begin{tabular}{>{\centering}p{4.3mm}>{\centering}p{4.3mm}>{\centering}p{4.3mm}>{\centering}p{4.3mm}>{\centering}p{4.3mm}>{\centering}p{4.3mm}>{\centering}p{4.3mm}>{\centering}p{4.3mm}>{\centering}p{4.3mm}>{\centering}p{4mm}>{\centering}p{4mm}>{\centering}p{4mm}>{\centering}p{4mm}>{\centering}p{4mm}>{\centering}p{4mm}>{\centering}p{4mm}>{\centering}p{4mm}>{\centering}p{4mm}>{\centering}p{2.3mm}>{\centering}p{1.3mm}}
\hline 
\noalign{\vskip0.8mm}
{\scriptsize{}$2^{19}$} & {\scriptsize{}$2^{18}$} & {\scriptsize{}$2^{17}$} & {\scriptsize{}$2^{16}$} & {\scriptsize{}$2^{15}$} & {\scriptsize{}$2^{14}$} & {\scriptsize{}$2^{13}$} & {\scriptsize{}$2^{12}$} & \textbf{\scriptsize{}$2^{11}$} & \textbf{\scriptsize{}$2^{10}$} & \textbf{\scriptsize{}$2^{9}$} & \textbf{\scriptsize{}$2^{8}$} & \textbf{\scriptsize{}$2^{7}$} & \textbf{\scriptsize{}$2^{6}$} & \textbf{\scriptsize{}$2^{5}$} & \textbf{\scriptsize{}$2^{4}$} & \textbf{\scriptsize{}$2^{3}$} & \textbf{\scriptsize{}$2^{2}$} & \textbf{\scriptsize{}$2^{1}$} & \textbf{\scriptsize{}$2^{0}$}\tabularnewline[0.8mm]
\hline 
\hline 
 &  &  &  &  &  &  &  &  &  & {\scriptsize{}1} & {\scriptsize{}$p_{8}$} & {\scriptsize{}$p_{7}$} & {\scriptsize{}$p_{6}$} & {\scriptsize{}$p_{5}$} & {\scriptsize{}$p_{4}$} & {\scriptsize{}$p_{3}$} & {\scriptsize{}$p_{2}$} & {\scriptsize{}$p_{1}$} & {\scriptsize{}1}\tabularnewline
 &  &  &  &  &  &  &  &  &  & {\scriptsize{}1} & {\scriptsize{}$q_{8}$} & {\scriptsize{}$q_{7}$} & {\scriptsize{}$q_{6}$} & {\scriptsize{}$q_{5}$} & {\scriptsize{}$q_{4}$} & {\scriptsize{}$q_{3}$} & {\scriptsize{}$q_{2}$} & {\scriptsize{}$q_{1}$} & {\scriptsize{}1}\tabularnewline[2mm]
\hline 
\noalign{\vskip0.6mm}
 &  &  &  &  &  &  &  &  &  & {\scriptsize{}1} & {\scriptsize{}$p_{8}$} & {\scriptsize{}$p_{7}$} & {\scriptsize{}$p_{6}$} & {\scriptsize{}$p_{5}$} & {\scriptsize{}$p_{4}$} & {\scriptsize{}$p_{3}$} & {\scriptsize{}$p_{2}$} & {\scriptsize{}$p_{1}$} & {\scriptsize{}1}\tabularnewline
 &  &  &  &  &  &  &  &  & {\scriptsize{}$q_{1}$} & {\scriptsize{}$p_{8}q_{1}$} & {\scriptsize{}$p_{7}q_{1}$} & {\scriptsize{}$p_{6}q_{1}$} & {\scriptsize{}$p_{5}q_{1}$} & {\scriptsize{}$p_{4}q_{1}$} & {\scriptsize{}$p_{3}q_{1}$} & {\scriptsize{}$p_{2}q_{1}$} & {\scriptsize{}$p_{1}q_{1}$} & {\scriptsize{}$q_{1}$} & \tabularnewline
 &  &  &  &  &  &  &  & {\scriptsize{}$q_{2}$} & {\scriptsize{}$p_{8}q_{2}$} & {\scriptsize{}$p_{7}q_{2}$} & {\scriptsize{}$p_{6}q_{2}$} & {\scriptsize{}$p_{5}q_{2}$} & {\scriptsize{}$p_{4}q_{2}$} & {\scriptsize{}$p_{3}q_{2}$} & {\scriptsize{}$p_{2}q_{2}$} & {\scriptsize{}$p_{1}q_{2}$} & {\scriptsize{}$q_{2}$} &  & \tabularnewline
 &  &  &  &  &  &  & {\scriptsize{}$q_{3}$} & {\scriptsize{}$p_{8}q_{3}$} & {\scriptsize{}$p_{7}q_{3}$} & {\scriptsize{}$p_{6}q_{3}$} & {\scriptsize{}$p_{5}q_{3}$} & {\scriptsize{}$p_{4}q_{3}$} & {\scriptsize{}$p_{3}q_{3}$} & {\scriptsize{}$p_{2}q_{3}$} & {\scriptsize{}$p_{1}q_{3}$} & {\scriptsize{}$q_{3}$} &  &  & \tabularnewline
 &  &  &  &  &  & {\scriptsize{}$q_{4}$} & {\scriptsize{}$p_{8}q_{4}$} & {\scriptsize{}$p_{7}q_{4}$} & {\scriptsize{}$p_{6}q_{4}$} & {\scriptsize{}$p_{5}q_{4}$} & {\scriptsize{}$p_{4}q_{4}$} & {\scriptsize{}$p_{3}q_{4}$} & {\scriptsize{}$p_{2}q_{4}$} & {\scriptsize{}$p_{1}q_{4}$} & {\scriptsize{}$q_{4}$} &  &  &  & \tabularnewline
 &  &  &  &  & {\scriptsize{}$q_{5}$} & {\scriptsize{}$p_{8}q_{5}$} & {\scriptsize{}$p_{7}q_{5}$} & {\scriptsize{}$p_{6}q_{5}$} & {\scriptsize{}$p_{5}q_{5}$} & {\scriptsize{}$p_{4}q_{5}$} & {\scriptsize{}$p_{3}q_{5}$} & {\scriptsize{}$p_{2}q_{5}$} & {\scriptsize{}$p_{1}q_{5}$} & {\scriptsize{}$q_{5}$} &  &  &  &  & \tabularnewline
 &  &  &  & {\scriptsize{}$q_{6}$} & {\scriptsize{}$p_{8}q_{6}$} & {\scriptsize{}$p_{7}q_{6}$} & {\scriptsize{}$p_{6}q_{6}$} & {\scriptsize{}$p_{5}q_{6}$} & {\scriptsize{}$p_{4}q_{6}$} & {\scriptsize{}$p_{3}q_{6}$} & {\scriptsize{}$p_{2}q_{6}$} & {\scriptsize{}$p_{1}q_{6}$} & {\scriptsize{}$q_{6}$} &  &  &  &  &  & \tabularnewline
 &  &  & {\scriptsize{}$q_{7}$} & {\scriptsize{}$p_{8}q_{7}$} & {\scriptsize{}$p_{7}q_{7}$} & {\scriptsize{}$p_{6}q_{7}$} & {\scriptsize{}$p_{5}q_{7}$} & {\scriptsize{}$p_{4}q_{7}$} & {\scriptsize{}$p_{3}q_{7}$} & {\scriptsize{}$p_{2}q_{7}$} & {\scriptsize{}$p_{1}q_{7}$} & {\scriptsize{}$q_{7}$} &  &  &  &  &  &  & \tabularnewline
 &  & {\scriptsize{}$q_{8}$} & {\scriptsize{}$p_{8}q_{8}$} & {\scriptsize{}$p_{7}q_{8}$} & {\scriptsize{}$p_{6}q_{8}$} & {\scriptsize{}$p_{5}q_{8}$} & {\scriptsize{}$p_{4}q_{8}$} & {\scriptsize{}$p_{3}q_{8}$} & {\scriptsize{}$p_{2}q_{8}$} & {\scriptsize{}$p_{1}q_{8}$} & {\scriptsize{}$q_{8}$} &  &  &  &  &  &  &  & \tabularnewline
 & {\scriptsize{}1} & {\scriptsize{}$p_{8}$} & {\scriptsize{}$p_{7}$} & {\scriptsize{}$p_{6}$} & {\scriptsize{}$p_{5}$} & {\scriptsize{}$p_{4}$} & {\scriptsize{}$p_{3}$} & {\scriptsize{}$p_{2}$} & {\scriptsize{}$p_{1}$} & {\scriptsize{}1} &  &  &  &  &  &  &  &  & \tabularnewline
{\scriptsize{}$z_{18,19}$} & {\scriptsize{}$z_{17,18}$} & {\scriptsize{}$z_{16,17}$} & {\scriptsize{}$z_{15,16}$} & {\scriptsize{}$z_{14,15}$} & {\scriptsize{}$z_{13,14}$} & {\scriptsize{}$z_{12,13}$} & {\scriptsize{}$z_{11,12}$} & {\scriptsize{}$z_{10,11}$} & {\scriptsize{}$z_{9,10}$} & {\scriptsize{}$z_{8,9}$} & {\scriptsize{}$z_{7,8}$} & {\scriptsize{}$z_{6,7}$} & {\scriptsize{}$z_{5,6}$} & {\scriptsize{}$z_{4,5}$} & {\scriptsize{}$z_{3,4}$} & {\scriptsize{}$z_{2,3}$} & {\scriptsize{}$z_{1,2}$} &  & \tabularnewline
{\scriptsize{}$z_{17,19}$} & {\scriptsize{}$z_{16,18}$} & {\scriptsize{}$z_{15,17}$} & {\scriptsize{}$z_{14,16}$} & {\scriptsize{}$z_{13,15}$} & {\scriptsize{}$z_{12,14}$} & {\scriptsize{}$z_{11,13}$} & {\scriptsize{}$z_{10,12}$} & {\scriptsize{}$z_{9,11}$} & {\scriptsize{}$z_{8,10}$} & {\scriptsize{}$z_{7,9}$} & {\scriptsize{}$z_{6,8}$} & {\scriptsize{}$z_{5,7}$} & {\scriptsize{}$z_{4,6}$} & {\scriptsize{}$z_{3,5}$} & {\scriptsize{}$z_{2,4}$} &  &  &  & \tabularnewline
 & {\scriptsize{}$z_{15,18}$} & {\scriptsize{}$z_{14,17}$} & {\scriptsize{}$z_{13,16}$} & {\scriptsize{}$z_{12,15}$} & {\scriptsize{}$z_{11,14}$} & {\scriptsize{}$z_{10,13}$} & {\scriptsize{}$z_{9,12}$} & {\scriptsize{}$z_{8,11}$} & {\scriptsize{}$z_{7,10}$} & {\scriptsize{}$z_{6,9}$} & {\scriptsize{}$z_{5,8}$} & {\scriptsize{}$z_{4,7}$} &  &  &  &  &  &  & \tabularnewline
 &  &  &  & {\scriptsize{}$z_{11,15}$} & {\scriptsize{}$z_{10,14}$} & {\scriptsize{}$z_{9,13}$} & {\scriptsize{}$z_{8,12}$} &  &  &  &  &  &  &  &  &  &  &  & \tabularnewline[0.6mm]
\hline 
\noalign{\vskip0.6mm}
{\scriptsize{}0} & {\scriptsize{}1} & {\scriptsize{}0} & {\scriptsize{}0} & {\scriptsize{}0} & {\scriptsize{}1} & {\scriptsize{}1} & {\scriptsize{}1} & {\scriptsize{}0} & {\scriptsize{}0} & {\scriptsize{}0} & {\scriptsize{}1} & {\scriptsize{}1} & {\scriptsize{}1} & {\scriptsize{}1} & {\scriptsize{}0} & {\scriptsize{}1} & {\scriptsize{}1} & {\scriptsize{}1} & {\scriptsize{}1}\tabularnewline[0.6mm]
\hline 
\end{tabular}
\end{table*}

\section*{Quantum factorization of larger numbers with the same 4-qubit Hamiltonian}
\vspace{-2mm}

Let us see what the Hamiltonian looks like for the factorization of
a larger number, using 56153 as an example. The multiplication table
in binary is given in Table \ref{tab:MutiplicationFor}. 

The equations obtained from adding the columns in the multiplication
table are then:

%\vspace{-15mm}
%\vspace{-15mm}
\textcolor{black}{\small{}
\begin{eqnarray*}
\mbox{\ensuremath{p_{1}+q_{1}}} & \negmedspace\negmedspace\negmedspace= & \negmedspace\negmedspace\negmedspace0+2z_{1,2}\\
\mbox{\ensuremath{p_{2}+p_{1}q_{1}+q_{2}+z_{1,2}}} & \negmedspace\negmedspace\negmedspace= & \mbox{\negmedspace\negmedspace\negmedspace\ensuremath{0+2z_{2,3}+4z_{2,4}}}\\
\mbox{\ensuremath{p_{3}+p_{2}q_{1}+p_{1}q_{2}+q_{3}+z_{2,3}}} & \negmedspace\negmedspace\negmedspace= & \negmedspace\negmedspace\negmedspace1+2z_{3,4}+4z_{3,5}\\
 & \negmedspace\negmedspace\negmedspace\vdots & \negmedspace\negmedspace\negmedspace\\
\mbox{\ensuremath{q_{6}+p_{6}+z_{12,13}+z_{11,13}+z_{10,13}}} & \negmedspace\negmedspace\negmedspace= & \negmedspace\negmedspace\negmedspace\mbox{\ensuremath{0+2z_{13,14}+4z_{13,15}}}\\
\mbox{\ensuremath{1+z_{13,14}+z_{12,14}+z_{11,14}}} & \negmedspace\negmedspace\negmedspace= & \mbox{\ensuremath{\negmedspace\negmedspace\negmedspace1+2z_{14,15}}}\\
z_{14,15}+z_{13,15} & \negmedspace\negmedspace\negmedspace= & \negmedspace\negmedspace\negmedspace\mbox{\ensuremath{1}}
\end{eqnarray*}
}{\small \par}

\vspace{-3mm}

\noindent and when the simplification rules are applied automatically
by a computer program, most $p_{i}$ and $q_{i}$ are already determined,
and the result is this set of equations:

\vspace{-5mm}

\begin{eqnarray}
p_{3}+q_{3} & = & 1\label{eq:XuEquations1-1}\\
p_{4}+q_{4} & = & 1\label{eq:XuEquations2-1}\\
p_{4}q_{3}+p_{3}q_{4} & = & 1.\label{eq:XuEquations3-1}
\end{eqnarray}

\vspace{0mm}

Notice how these have precisely the same form as the equations in
the factorization of 143, except with different variables. Therefore
the Hamiltonian is also the same, except the qubits $p_{a}$, $p_{b}$,
$q_{a}$, $q_{b}$ appearing represent different positions in the
corresponding binary strings representing $p$ and $q$. Other numbers
that we have discovered reduce to these same equations include 3599,
11663, and 56153.

In fact, it turns out that the product of any two numbers differing
at only 2 bits will lead to the equations:

\vspace{-8mm}

\begin{eqnarray}
p_{a}+q_{a} & = & x\label{eq:generalXuEquation1}\\
p_{b}+q_{b} & = & y\label{eq:generalXuEquation2}\\
p_{b}q_{a}+p_{a}q_{b} & = & z,\label{eq:generalXuEquation3}
\end{eqnarray}

\vspace{-1mm}
\noindent where the subscripts $a$ and $b$ correspond to the two bit-positions
that differ, and the right-side variables $\{x,y,z\}$ can each be
0 or 1 depending on the number being factored. However, unless we
know in advance that the factors will differ at two bits, this reduction
will not allow us to crack big RSA codes. Furthermore, Eqs. \ref{eq:generalXuEquation1}-\ref{eq:generalXuEquation3}
can easily be solved by a classical computer, since there are only
4 variables, and therefore solving only involves at most $2^{4}=16$
queries.\vspace{-5mm}

\section*{Cases that cannot be solved so easily on a classical computer
}

\begin{table*}
\protect\caption{Multiplication table for 175 \label{tab:MutiplicationFor175}}

{\scriptsize{}}%
\begin{tabular}{>{\centering}p{2mm}ccccccccc}
\hline 
 & \textbf{\scriptsize{}$2^{8}$} & \textbf{\scriptsize{}$2^{7}$} & \textbf{\scriptsize{}$2^{6}$} & \textbf{\scriptsize{}$2^{5}$} & \textbf{\scriptsize{}$2^{4}$} & \textbf{\scriptsize{}$2^{3}$} & \textbf{\scriptsize{}$2^{2}$} & \textbf{\scriptsize{}$2^{1}$} & \textbf{\scriptsize{}$2^{0}$}\tabularnewline
\hline 
\hline 
{\scriptsize{}$p$} &  &  &  &  &  &  & {\scriptsize{}1} & {\scriptsize{}$p_{1}$} & {\scriptsize{}1}\tabularnewline
{\scriptsize{}$q$} &  &  &  &  &  &  & {\scriptsize{}1} & {\scriptsize{}$q_{1}$} & {\scriptsize{}1}\tabularnewline
\hline 
 &  &  &  &  &  &  & {\scriptsize{}1} & {\scriptsize{}$p_{1}$} & {\scriptsize{}1}\tabularnewline
 &  &  &  &  &  & {\scriptsize{}$q_{1}$} & {\scriptsize{}$p_{1}q_{1}$} & {\scriptsize{}$q_{1}$} & \tabularnewline
 &  &  &  &  & {\scriptsize{}1} & {\scriptsize{}$p_{1}$} & {\scriptsize{}1} &  & \tabularnewline
\hline 
 &  &  &  & {\scriptsize{}$z_{45}$} & {\scriptsize{}$z_{34}$} & {\scriptsize{}$z_{23}$} & {\scriptsize{}$z_{12}$} &  & \tabularnewline
 &  &  &  & {\scriptsize{}$z_{35}$} & {\scriptsize{}$z_{24}$} &  &  &  & \tabularnewline
\hline 
 &  &  &  & {\scriptsize{}$z_{45}+z_{35}$} & {\scriptsize{}$1+z_{34}+z_{24}$} & {\scriptsize{}$q_{1}+p_{1}+z_{23}$} & {\scriptsize{}$2+p_{1}q_{1}+z_{12}$} & {\scriptsize{}$p_{1}+q_{1}$} & {\scriptsize{}1}\tabularnewline
{\scriptsize{}$r$} &  &  &  &  &  &  & {\scriptsize{}1} & {\scriptsize{}$r_{1}$} & {\scriptsize{}1}\tabularnewline
\hline 
 &  &  &  & {\scriptsize{}$z_{45}+z_{35}$} & {\scriptsize{}$1+z_{34}+z_{24}$} & {\scriptsize{}$q_{1}+p_{1}+z_{23}$} & {\scriptsize{}$2+p_{1}q_{1}+z_{12}$} & {\scriptsize{}$p_{1}+q_{1}$} & {\scriptsize{}1}\tabularnewline
 &  &  & {\scriptsize{}$r_{1}(z_{45}+z_{35})$} & {\scriptsize{}$r_{1}(1+z_{34}+z_{24})$} & {\scriptsize{}$r_{1}(q_{1}+p_{1}+z_{23})$} & {\scriptsize{}$r_{1}(2+p_{1}q_{1}+z_{12})$} & {\scriptsize{}$r_{1}(p_{1}+q_{1})$} & {\scriptsize{}$r_{1}$} & \tabularnewline
 &  & {\scriptsize{}$z_{45}+z_{35}$} & {\scriptsize{}$1+z_{34}+z_{24}$} & {\scriptsize{}$q_{1}+p_{1}+z_{23}$} & {\scriptsize{}$2+p_{1}q_{1}+z_{12}$} & {\scriptsize{}$p_{1}+q_{1}$} & {\scriptsize{}1} &  & \tabularnewline
\hline 
\noalign{\vskip0.6mm}
 & {\scriptsize{}$z_{78}^{'}$} & {\scriptsize{}$z_{67}^{'}$} & {\scriptsize{}$z_{56}^{'}$} & {\scriptsize{}$z_{45}^{'}$} & {\scriptsize{}$z_{34}^{'}$} & {\scriptsize{}$z_{23}^{'}$} & {\scriptsize{}$z_{12}^{'}$} &  & \tabularnewline
 & {\scriptsize{}$z_{68}^{'}$} & {\scriptsize{}$z_{57}^{'}$} & {\scriptsize{}$z_{46}^{'}$} & {\scriptsize{}$z_{35}^{'}$} & {\scriptsize{}$z_{24}^{'}$} &  &  &  & \tabularnewline[0.6mm]
\hline 
 & {\scriptsize{}0} & {\scriptsize{}1} & {\scriptsize{}0} & {\scriptsize{}1} & {\scriptsize{}0} & {\scriptsize{}1} & {\scriptsize{}1} & {\scriptsize{}1} & {\scriptsize{}1}\tabularnewline
\hline 
\end{tabular}{\scriptsize \par}

\vspace{0mm}
\end{table*}

\vspace{-2mm}

The real advantage of solving such equations as Eqs. \ref{eq:generalXuEquation1}-\ref{eq:generalXuEquation3}
via finding the ground state of an appropriate Hamiltonian, is realized
when the final set of reduced equations has many more variables.
Considering for example a case where the final equations have 512
unknowns, which is the number of qubits in the D-Wave Two, a brute
force ``guess and check'' strategy for solving the equation system
would require at most $2^{512}=10^{154}$ queries. If a trillion queries
could be made per second, this would amount to $\approx10^{123}$
times the age of the universe (clearly for numbers this large, classical
factorization algorithms alternative to solving the discrete minimization
problem would vastly outperform the ``brute force'' strategy, but
the best such classical algorithm is the General Number Field Sieve
whose computational complexity also contains an exponential). 

The question then is, which cases will reduce to a set of equations
with a large number of unknown variables (ie., which cases will be
able to exploit the use of more qubits). While we have noticed the
pattern that whenever the factors of a semiprime differ at two bit-positions,
the minimization problem contains 4 unknown variables (and the Hamiltonian
contains 4 qubits); we have also noticed that when the factors of
a semiprime differ at three bit-positions, the minimization problem
contains 6 unknown variables (see for example Table \textcolor{green}{\ref{tab:10x10291311}}
and \ref{eq:6qubitEquations1}-\ref{eq:6qubitEquatons6}). In fact,
the semiprimes which exploit the most power from the quantum computer
will be those whose factors differ in the largest number of possible
digits. \textcolor{black}{For an example of this see Table}\textcolor{green}{{}
\ref{tab:10x10291311}.}

From Table \ref{tab:10x10291311} the following equations are ultimately
derived:

\vspace{-10mm}

\begin{eqnarray}
p_{1}+q_{1} & = & 1\label{eq:6qubitEquations1}\\
p_{2}+q_{2} & = & 1\\
p_{5}+q_{5} & = & 1\\
p_{1}q_{2}+q_{1}p_{2} & = & 1\\
p_{2}q_{5}+q_{2}p_{5} & = & 0\\
p_{5}q_{1}+q_{5}p_{1} & = & 1,\label{eq:6qubitEquatons6}
\end{eqnarray}

\vspace{-2mm}
\noindent whose solution is encoded in the ground state of the following Hamiltonian
involving at most 3-qubit interactions:

\end{multicols}

\vspace{-6mm}

{\scriptsize{}
\begin{eqnarray}
H & \negmedspace\negmedspace\negmedspace\negmedspace= & \negmedspace\negmedspace\negmedspace\negmedspace(p_{1}+q_{1}-1)^{2}+(p_{2}+q_{2}-1)^{2}+(p_{5}+q_{5}-1)^{2}+(p_{1}q_{2}+q_{1}p_{2}-1)^{2}+p_{2}q_{5}+q_{2}p_{5}+(p_{5}q_{1}+q_{5}p_{1}-1)^{2}\\
 & \negmedspace\negmedspace\negmedspace\negmedspace= & \negmedspace\negmedspace\negmedspace\negmedspace9\negthinspace+\negthinspace2p_{2}p_{1}q_{1}\negthinspace+\negthinspace2p_{2}q_{2}q_{1}\negthinspace+\negthinspace2p_{5}p_{1}q_{1}\negthinspace+\negthinspace2p_{5}q_{5}q_{1}\negthinspace+\negthinspace2p_{1}q_{1}\negthinspace+\negthinspace2p_{2}q_{2}\negthinspace+\negthinspace2p_{5}q_{5}\negthinspace-\negthinspace3p_{2}q_{1}\negthinspace+\negthinspace p_{1}q_{2}\negthinspace-\negthinspace3p_{5}q_{1}\negthinspace+\negthinspace p_{1}q_{5}\negthinspace-\negthinspace5p_{1}-\negthinspace p_{2}\negthinspace-\negthinspace q_{1}\negthinspace-\negthinspace3q_{2}\negthinspace-\negthinspace3q_{5}\label{eq:6qubitHamiltonian}\\
a_{i} & \negmedspace\negmedspace\negmedspace\negmedspace= & \negmedspace\negmedspace\negmedspace\negmedspace\frac{1}{2}\left(1-\sigma_{z}^{(i)}\right)^{2}.
\end{eqnarray}
}{\footnotesize{} }\vspace{-8mm}

\begin{table*}
\protect\caption{Quantum factorization records\label{tab:quantumFactorizationRecords}}

\centering{}%
\begin{tabular}{>{\centering}m{18mm}>{\centering}m{25mm}>{\centering}m{20mm}>{\centering}m{23mm}>{\centering}m{24mm}>{\centering}m{30mm}}
\hline 
Number & \# of factors & \# of qubits needed & Algorithm & Year implemented & Implemented without prior knowledge of solution\tabularnewline
\hline 
\noalign{\vskip2mm}
15 & 2 & 8 & Shor  & 2001 \cite{Vandersypen2001} & \xmark\tabularnewline
 & 2 & 8 & Shor  & 2007 \cite{Lanyon2007} & \xmark\tabularnewline
 & 2 & 8 & Shor  & 2007 \cite{Lanyon2007} & \xmark\tabularnewline
 & 2 & 8 & Shor  & 2009 \cite{Politi2009} & \xmark\tabularnewline
 & 2 & 8 & Shor  & 2012 \cite{Martin-Lopez2012} & \xmark\tabularnewline
21 & 2 & 10 & Shor  & 2012 \cite{Lucero2012} & \xmark\tabularnewline
143 & 2 & 4 & minimization & 2012 \cite{Xu2012} & \cmark\tabularnewline
56153 & 2 & 4 & minimization & 2012 \cite{Xu2012} & \cmark\tabularnewline[2mm]
\hline 
\noalign{\vskip2mm}
291311 & 2 & 6 & minimization & not yet  & \cmark\tabularnewline
175 & 3 & 3 & minimization & not yet & \cmark\tabularnewline
\end{tabular}
\end{table*}

\begin{multicols}{2}

Finally, it has been shown by Schutzhold and Schaller \cite{Schaller2010}
that the Hamiltonians whose ground states encode the solutions to
the factorization problem for semiprimes will always have at most
3-body interactions between the qubits (as seen for example in the
Hamiltonians of Eqs.\ref{eq:4qubitHamiltonian} \& \ref{eq:6qubitHamiltonian}).
Such Hamiltonians are relatively simple to implement in experiments
\cite{Schaller2010,Xu2012}. The only cases that will involve more
``difficult'' to implement Hamiltonians (in terms of requiring many-body
interactions between qubits) will be where the integer to be factored
is the product of more than 2 numbers (the more factors, the higher-body
interactions required). The construction of the factorization table
when there are more than 2 factors is a bit more complicated, so we
demonstrate the factorization of $175$ below. Interestingly, until
now, the only numbers for which successful quantum factorization has
been demonstrated, are integers with only 2 factors.  

For the factorization of 175, Table \ref{tab:MutiplicationFor175}
leads t\textcolor{black}{o the equations:}

%\textcolor{green}{}
\vspace{-5mm}

\begin{flushleft}
\textcolor{black}{\footnotesize{}
\begin{eqnarray}
\mbox{\ensuremath{p_{1}\negmedspace+\negmedspace q_{1}\negmedspace+\negmedspace r_{1}}} & \negmedspace\negmedspace\negmedspace\negmedspace= & \mbox{\negmedspace\negmedspace\ensuremath{\negmedspace\negmedspace1\negmedspace+\negmedspace2z_{12}^{'}}}\label{eq:triprimeEqnsStart}\\
\mbox{\ensuremath{3+p_{1}q_{1}\negmedspace+\negmedspace z_{12}\negmedspace+\negmedspace r_{1}(p_{1}\negmedspace+\negmedspace q_{1})\negmedspace+\negmedspace z_{12}^{'}}} & \negmedspace\negmedspace\negmedspace\negmedspace= & \mbox{\negmedspace\negmedspace\negmedspace\negmedspace\ensuremath{1\negmedspace+\negmedspace2z_{23}^{'}\negmedspace+\negmedspace4z_{24}^{'}\qquad}}\\
\mbox{\ensuremath{z_{23}\negmedspace+\negmedspace r_{1}(2\negmedspace+p_{1}q_{1}\negmedspace+z_{12})\negmedspace+\negmedspace2p_{1}\negmedspace+\negmedspace2q_{1}\negmedspace+\negmedspace z_{23}^{'}}} & \negmedspace\negmedspace\negmedspace\negmedspace= & \negmedspace\negmedspace\negmedspace\negmedspace1\negmedspace+\negmedspace2z_{34}^{'}\negmedspace+\negmedspace4z_{35}^{'}\qquad\\
 & \negmedspace\negmedspace\negmedspace\negmedspace\vdots & \negmedspace\negmedspace\negmedspace\negmedspace\nonumber \\
\mbox{\ensuremath{r_{1}(z_{45}\negmedspace+\negmedspace z_{35})\negmedspace+\negmedspace1\negmedspace+\negmedspace z_{34}\negmedspace+\negmedspace z_{24}\negmedspace+\negmedspace z_{56}^{'}\negmedspace+\negmedspace z_{46}^{'}}} & \negmedspace\negmedspace\negmedspace\negmedspace= & \mbox{\negmedspace\negmedspace\negmedspace\negmedspace\ensuremath{0\negmedspace+\negmedspace2z_{67}^{'}\negmedspace+\negmedspace4z_{68}^{'}}}\qquad\\
\mbox{\ensuremath{z_{45}\negmedspace+\negmedspace z_{35}\negmedspace+\negmedspace z_{67}^{'}\negmedspace+\negmedspace z_{57}^{'}}} & \negmedspace\negmedspace\negmedspace\negmedspace= & \negmedspace\negmedspace\mbox{\negmedspace\negmedspace\ensuremath{1\negmedspace+\negmedspace2z_{78}^{'}}}\\
\mbox{\ensuremath{z_{78}^{'}\negmedspace+\negmedspace z_{68}^{'}}} & \negmedspace\negmedspace\negmedspace\negmedspace= & \negmedspace\negmedspace\negmedspace\negmedspace0\label{eq:triprimeEqnsEnd}
\end{eqnarray}
}
\par\end{flushleft}{\footnotesize \par}

\vspace{2mm}

\begin{flushleft}
\vspace{-20mm}
which ultimately reduce\textcolor{black}{{} to: }\textcolor{red}{\scriptsize{}}
\par\end{flushleft}{\scriptsize \par}

\begin{flushleft}

\vspace{-7mm}

{\small 
\textcolor{black}{
\begin{eqnarray}
p_{1}+q_{1}+r_{1} & = & 1\label{eq:triprimeReducedEqn}\\
p_{1}q_{1}+q_{1}r_{1}+p_{1}r_{1} & = & 0.\label{eq:triprimeReducedEqn2}
\end{eqnarray}
}}
\par\end{flushleft}

\vspace{-4mm}

\begin{flushleft}
\vspace{-12mm}
The Hamiltonian whose ground state encodes the solution to Eqs. \ref{eq:triprimeReducedEqn}-\ref{eq:triprimeReducedEqn2}
is then:
\par\end{flushleft}

\vspace{-7mm}

{\small{}
\begin{eqnarray*}
H & \negmedspace\negmedspace\negmedspace\negmedspace=\negmedspace\negmedspace\negmedspace\negmedspace & (p_{1}+q_{1}+r_{1}-1)^{2}+(p_{1}q_{1}+q_{1}r_{1}+p_{1}r_{1})^{2}\\
 & \negmedspace\negmedspace\negmedspace\negmedspace=\negmedspace\negmedspace\negmedspace\negmedspace & 1+2p_{1}q_{1}+6p_{1}q_{1}r_{1}+2p_{1}r_{1}+2q_{1}r_{1}-p_{1}-q_{1}-r_{1}\qquad\\
 & \negmedspace\negmedspace\negmedspace\negmedspace\negmedspace\negmedspace\negmedspace\negmedspace\\
 & \negmedspace\negmedspace\negmedspace\negmedspace=\negmedspace\negmedspace\negmedspace\negmedspace & \begin{pmatrix}1 & 0 & 0 & 0 & 0 & 0 & 0 & 0\\
0 & \mathbf{0} & 0 & 0 & 0 & 0 & 0 & 0\\
0 & 0 & \mathbf{0} & 0 & 0 & 0 & 0 & 0\\
0 & 0 & 0 & 2 & 0 & 0 & 0 & 0\\
0 & 0 & 0 & 0 & \mathbf{0} & 0 & 0 & 0\\
0 & 0 & 0 & 0 & 0 & 2 & 0 & 0\\
0 & 0 & 0 & 0 & 0 & 0 & 2 & 0\\
0 & 0 & 0 & 0 & 0 & 0 & 0 & 7
\end{pmatrix}.
\end{eqnarray*}
}{\small \par}

\vspace{2mm}

The lowest eigenvalues correspond to $|1\rangle=|001\rangle$, $|2\rangle=|010\rangle$,
and $|4\rangle=|100\rangle$, which (respectively) correspond to:

\vspace{-2mm}

\begin{eqnarray}
(p_{1},q_{1},r_{1})\negmedspace\negmedspace\negmedspace & =\negmedspace\negmedspace\negmedspace & (0,0,1)\rightarrow p=5,\, q=5,\, r=7,\qquad\\
(p_{1},q_{1},r_{1})\negmedspace\negmedspace\negmedspace & =\negmedspace\negmedspace\negmedspace & (0,1,0)\rightarrow p=5,\, q=7,\, r=5,\qquad\\
(p_{1},q_{1},r_{1})\negmedspace\negmedspace\negmedspace & =\negmedspace\negmedspace\negmedspace & (1,0,0)\rightarrow p=7,\, q=5,\, r=5.\qquad
\end{eqnarray}

\vspace{-2mm}

\section*{Conclusion}

\vspace{-2mm}

	We have shown that the NMR experiment of Xu \emph{et al.}\cite{Xu2012}
in 2012 factored an entire class of numbers, and not just the one
number that they reported (which was 143). The largest such number
that we found without using any prior knowledge of the solution to
the factorization problem was 56153. Since the experiment in \cite{Xu2012}
only involved 4 qubits, it could easily have been factored on a classical
computer as well. In order to exploit the true power of quantum mechanics
in this type of computation, finding the solution will need to make
use of more qubits. To this end, we have discussed a scheme for factoring
numbers via the same minimization technique, but where more qubits
are required to solve the discrete optimization problem. As an example,
we demonstrated how to factor 291311 with 6 qubits. To put this into
context, Table \ref{tab:quantumFactorizationRecords} shows all progress
until now in factoring numbers using quantum computers.

We further noted that the Hamiltonians involved in factoring numbers
via discrete minimization only involve 3-qubit interactions when the
numbers to factor are semiprimes. Such Hamiltonians are relatively
easy to construct (and it is noted that the best classical factorization
algorithms are the opposite: they find semiprimes most difficult).
Hamiltonians involving higher-body qubit interactions will only arise
for the factorization of numbers with more than two factors. To this
end, we have demonstrated the quantum factorization of the triprime
175 with 3 qubits. 

\bibliographystyle{IEEEtran}

\end{multicols}
\end{document}